\documentclass[twocolumn,showpacs,preprintnumbers,amsmath,amssymb]{revtex4}

\usepackage{graphicx}
\usepackage{dcolumn}
\usepackage{bm}

\begin{document}

\title{Four-body Faddeev-Yakubovsky calculation of
polarization transfer coefficient $\bm{K}_y^{y'}$
in the $^2$H$(\vec{\bm{d}}, \vec{\bm{p}})^3$H reaction at very low energies}

\author{Eizo Uzu}
\email{j-uzu@ed.noda.tus.ac.jp}
\affiliation{Department of Physics, Faculty of Science and Technology,
Tokyo University of Science,
2641 Noda, Chiba 278-8510, Japan}

\date{\today}

\begin{abstract}
The polarization transfer coefficient $K_y^{y'}$ for
the $^2$H$(\vec{d}, \vec{p})^3$H reaction at energies
$E_{\rm lab.} = 30$, 50, 70, and 90keV is calculated
using the four-body Faddeev-Yakubovsky equation.
We employ a separable type of the Paris potential.
The experimental data [Phys. Rev. C{\bf 64}, 047601 (2001)] agrees with
our solution at the scattering angle $\theta_{\rm c.m.}=0^\circ$,
and there is no discrepancy pointed out in that paper.
We found that the $K_y^{y'}$ is sensitive
to the incident deuteron energy at this angle.
\end{abstract}


\maketitle

Since Kulsrud et al. proposed the idea of a neutron-lean fusion reactor~\cite{KFVG},
low energy $^2$H$(d,p)^3$H and $^2$H$(d,n)^3$He reactions have attracted much
attention.
Subjects of discussions are mainly about suppression of the reaction
by the polarized beam and target deuterons in the same direction ($\vec{d}+\vec{d}$),
to that by the unpolarized ones ($d+d$).
Experimentally there is no direct data for the $\vec{d}+\vec{d}$ reaction,
however new data of the cross sections by the unpolarized $^2$H$(d,p)^3$H
and $^2$H$(d,n)^3$He reactions~\cite{LosAla} and the vector and tensor analyzing
powers by the polarized beam~\cite{Koeln,Tsuku} at several 10keV energies.

On the other hand there are many theoretical approaches to the suppression problem,
with models~\cite{HFsup,LZSsup,KIJTsup}
and analyses of the experimental data~\cite{LSsup,FABsup,ZLS2sup}.
But their solutions are all different each other depending the applied models or
analyses methods, and there is no conclusion.

To obtain more accurate solution, we employ the four-body Faddeev-Yakubovsky (FY)
equation~\cite{FaYa} which is able to calculate the system in microscopic way.
The FY theory is a direct extension of the three-body Faddeev theory to
the four-body system, and it includes all kinematics of four nucleons.
There are many vigorous studies in this field.
For recent example, the analyzing powers for 4N system~\cite{FON1,FON2,FON3},
and the cross section and low energy resonance for
$n^3$He elastic scattering~\cite{FON3,carbo} are reported.
And the FY equation is applied to the $^4$He bound state~\cite{KMD}
and hypernuclei~\cite{hyper1,hyper2}.
In regard to the $^4$He ground state, our numerical result for the binding energy
agrees with that by Fonseca~\cite{PrCom} within a limited
number of state channels and employing Yamaguchi potential.

In this decade, our investigations were mainly on the analyses
of the experimental data of the cross section and analyzing powers,
together with prediction of the suppression ratio~\cite{uzu1,uzu2,uzu3,uzu4,uzu5,uzu6}.
It began from Ref.~\cite{uzu1} and estimation of the suppression by
the invariant amplitude method~\cite{IAM}
is compared with our prediction using the FY equation in Ref.~\cite{uzu2}.
The final result employing the Yamaguchi-type nucleon-nucleon (NN) interaction
are presented in Ref.~\cite{uzu3,uzu4}.
And the first step using the realistic potential is discussed in Ref.~\cite{uzu5}.
We think that a study of the S-factor is one of our important subject~\cite{uzu6}
to present a data for the astrophysics.

Recently Katabuchi et al. measured the polarization transfer coefficient
$K_y^{y'}$ for the $^2$H$(\vec{d}, \vec{p})^3$H reaction~\cite{Kata}.
The incident deuteron laboratory energy ($E_d$) is 90keV,
however, the data included those for all energies between 0 - 90keV
with the energy loss in the target.
The acceptance angle of the detector was $\pm 6.5^{\circ}$ around
the scattering angle $0^\circ$ in the laboratory system.
The obtained data was $K_y^{y'} = 0.09 \pm 0.10$.

In the same paper, they reported numerical results of the $K_y^{y'}$
using the amplitudes determined by Lema\^{i}tre and Schieck~\cite{LSsup},
where the amplitudes are expressed as a production of
the Coulomb penetrability function and ``internal'' reaction amplitudes
in the NN interaction without the Coulomb force.
The ``internal'' amplitudes are treated as parameters independent of
the incident deuteron energy,
and fitted to many experimental data for the $d+d$ reactions.
The energy dependence is kept merely in the Coulomb penetrability function.
Obtained $K_y^{y'}$ has very small energy dependence in all scattering
angles, and the mean value over $E_{\rm lab.}$ is
different from the measured data by 1.5 standard deviation.
Thus they concluded that the other models should be compared with the data.

Complying to the discussion, we calculate the $K_y^{y'}$ at energies
$E_{\rm lab.} =$30, 50, 70, and 90keV using the FY equation
with following conditions.
As the NN interaction, the Ernst, Shakin, and Thaler separable expansion~\cite{EST}
of the Paris (PEST) potential~\cite{PEST} are employed for the following two-body
states: $^1$S$_0$ (3), $^3$S$_1$-$^3$D$_1$ (4), $^3$P$_0$ (2),
$^1$P$_1$ (2), $^3$P$_1$ (2), and $^3$P$_2$ (3).
The numbers in the braces are ranks for each states.
In the [3+1] sub-amplitude,
the three-body total spin and parity $j^\pi$ considered
are within the range $1/2^\pm$ to $7/2^\pm$.
For the [2+2] sub-amplitude, all possible quantum states,
which are automatically driven by two-body states, are taken into account.
As for the four-body total spin and parity states
~$J^\pi$ these are taken from $0^\pm$ to $4^\pm$ and the convergence is confirmed.

The [3+1] and [2+2] subamplitudes are expanded in separable form
adopting the energy dependent pole expansion (EDPE)~\cite{EDPE} method
with following modification.
It is convenient to describe an outline of the technique with the expressions
in Ref.~\cite{NLAH}.
The original EDPE method requires to solve the eigenequations (15)
with fixed energies $E_x$ and $E_y$ which are usually chosen below the $d+p+n$
threshold Energy.
We adopt the numerical results of the binding energies for the triton as $E_x$
and two deuterons as $E_y$ in our calculation, and they are used in all states
of the [3+1] and [2+2] subsystem.
The form factors and propagators are obtained from the third through sixth
equations in Eqs. (14) with required energy $z$ on the FY calculation.
In case for the mixed channels of the repulsive and attractive forces
or the multi-rank separable potentials in the 2-body subsystem,
Eqs. (15) have many complex solutions and the convergence of the FY calculation
become poor.
The modification intends to improve this difficulty.

First the 2-body propagator $\widetilde{\tau}$ is diagonalized
with a unitary matrix $U$ as
\begin{equation}
\widetilde{\tau}^D = U \widetilde{\tau} U^{-1}.
\end{equation}
Next all the positive diagonal elements are set by negative but the same absolute
values as
\begin{equation}
\widetilde{\tau}^{RD}_{jj} = - |\widetilde{\tau}^D_{jj}|.
\end{equation}
Thus the reduced propagator $\widetilde{\tau}^R$ is obtained
with $U$ as
\begin{equation}
\widetilde{\tau}^R = U^{-1} \widetilde{\tau}^{RD} U.
\end{equation}
and $\widetilde{\tau}^R$ is used in the eigenequations (15) except for $\widetilde{\tau}$.
In this case all the eigenvalues and eigenfunctions become real,
and the FY calculation keeps good convergence.
After obtaining the form factor, the [3+1] and [2+2] propagators are
calculated with the original $\widetilde{\tau}$.
Details of this technique will be described on the other paper.

As for the Coulomb force, exact calculation requires to distinguish the proton
and neutron, and to include the pure Coulomb potential between two protons.
However nobody is successful to carry out this method in the FY integral equation
within the momentum representation for the difficulty of the logarithmic singularity.
An approximation method is proposed in Ref.~\cite{LSsup} but we adopt a different
way of which basic idea is described as ``Traditional'' in Ref.~\cite{uzu6}.
In this approximation, the FY equation for 4-identical particles is solved
without Coulomb force, and obtained amplitudes are regarded as solutions for a
mixed channel 2-body problem where the two deuterons in the initial channel and
proton and triton in the final are treated as point particles.
The scattering amplitudes are described the operator of the T-matrix put between
the plain wave functions for the initial and final channels.
However in case for existence of the Coulomb repulsion, the Coulomb functions
ought to be applied except for the plain waves.
We modify this approximately with multiplying the amplitude by simple factors as
\begin{equation}
P'_{0} e^{\delta_{\ell'}} M_{\ell' \ell} P_{0} e^{\delta_{\ell}} \,,
\end{equation}
where $M_{\ell' \ell}$ is a reaction amplitude without Coulomb whose $\ell$ and $\ell'$
are the angular momentum in the initial and final channel, respectively.
$P_{0}$s and $\delta_{\ell}$s are the damping effects and phase shifts corrections,
respectively, where without and with prime indicates the initial and final channel,
respectively.
These factors are expressed as the ratio of the Coulomb wave function to the plain
wave which is applied the limiting value of the radial valuable $r$ to be 0 since
the range of the NN interaction is much smaller than that of the Coulomb force.

Thus we obtain $P_0$ as
\begin{equation}  
P_0 \equiv \lim_{r \to 0} \left| \frac{\psi_c({\bf r})}{\phi_k ({\bf r})} \right|
= \left| \exp (- \frac{1}{2} \pi \eta) \Gamma (1 + i \eta) \right|,
\label{eq:penet}
\end{equation}  
where $\psi_c({\bf r})$ and $\phi_k ({\bf r})$ are regular functions of
the Coulomb and plain waves, respectively.
$\eta= 0.1581/\sqrt{E_i \,}$ is the Sommerfeld parameter where $E_i$ is a
center of mass (c.m.) energy in the initial channel.
$P'_0$ has the same expression except for changing $\eta$ into
$\eta'= 0.1369 / \sqrt{E_f \,}$ where $E_f$ is a c.m. energy in the final channel.
It is easy to obtain the Gamow factor from $P_0^2$ in case for small energy which
is described for instance in Ref.~\cite{Schiff}.
This factor is cancelled in $K_y^{y'}$ due to independence of the partial waves.

The Coulomb phase shift $\delta_\ell$ is defined in the same manner as
\begin{equation}
\delta_\ell = \arg \Gamma(\ell + 1 + i \eta)
= \arg \left[ \lim_{r \to 0} \frac{\psi_\ell(r)}{\phi_\ell (r)} \right],
\end{equation}
where $\psi_\ell(r)$ and $\phi_\ell (r)$ are regular functions of the Coulomb
and plain waves, respectively, represented by the partial wave decomposition.
$\delta_\ell$ is represented in the same expression except for changing
$\ell$ and $\eta$ into $\ell'$ and $\eta'$, respectively.
It is first introduced in Ref.~\cite{CLBorg} for a two-body system
and applied it to the three-body system in Ref.~\cite{CLB3b}.

One may think that the modification will be improved by changes of the functions
$\psi_c({\bf r})$ and $\phi_k ({\bf r})$ in Eq. (\ref{eq:penet}) for
$\psi_\ell(r)$ and $\phi_\ell (r)$.
However in this case, the P-wave components in the initial channel are too much
enhanced and the results for the differential cross sections and analyzing powers
become worse.
Then we did not adopt it.

\begin{figure}[!t]
\includegraphics[width=80mm]{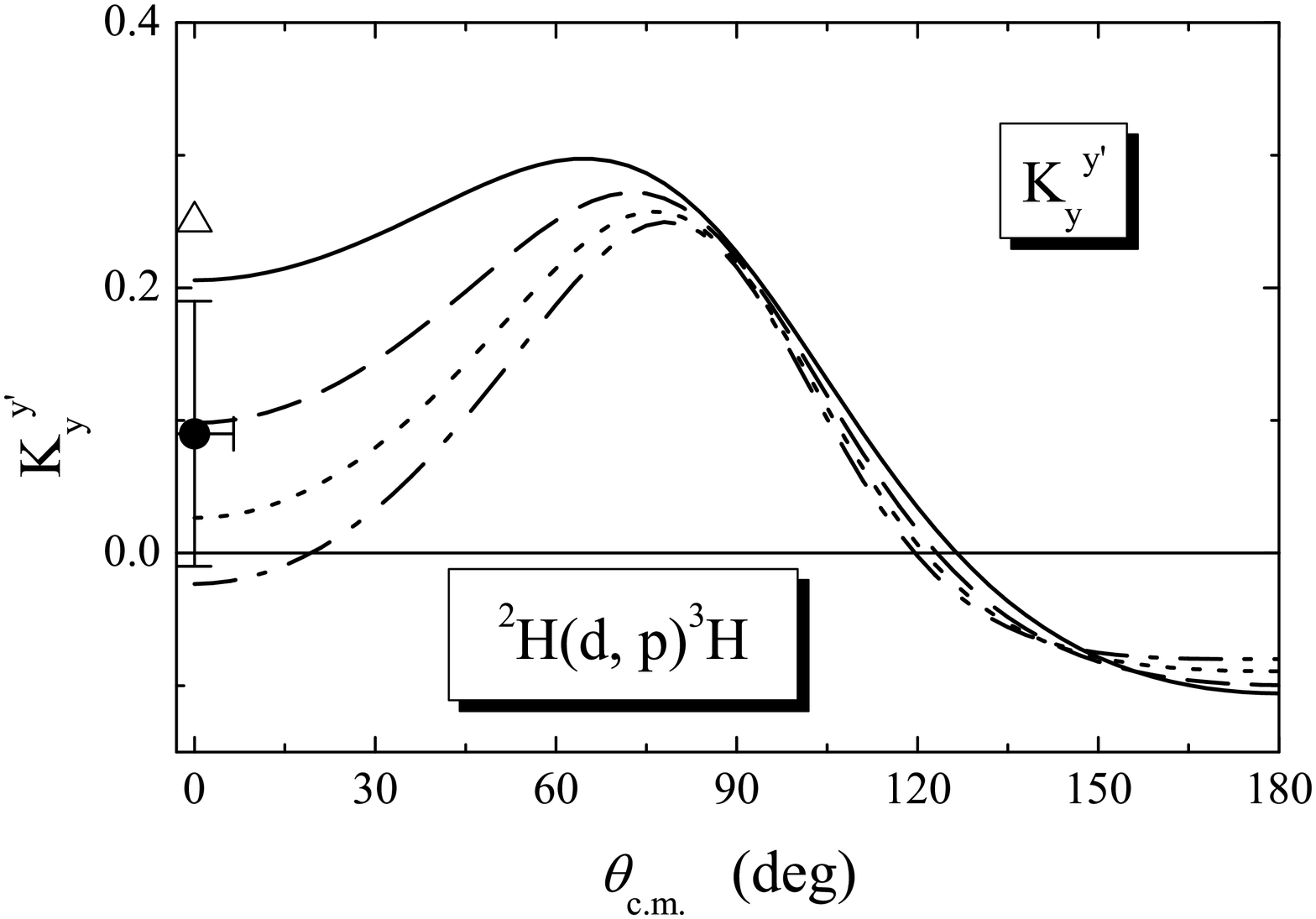}
\caption{The polarization transfer coefficient $K_y^{y'}$
at $E_{\rm lab.} =30$keV (solid line), 50keV (dashed line),
70keV (dotted line), and 90keV(dash-dotted line).
The measured data (solid circle) in Ref.~\cite{Kata} is compared
with our numerical results.
The triangle dot shows the mean value over the incident energies
of their calculational results in the same paper.
\label{fig:kyy}}
\end{figure}
\begin{figure}[!t]
\includegraphics[width=80mm]{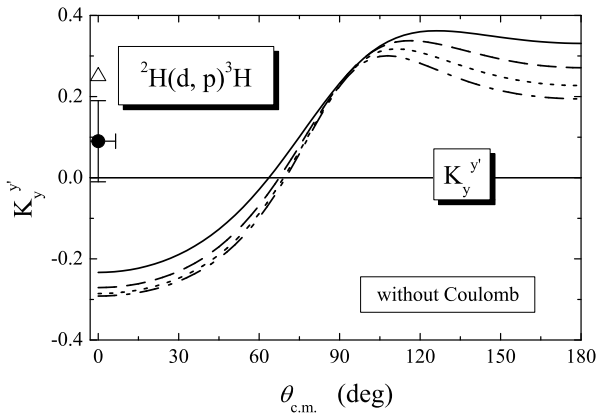}
\caption{Same as for FIG. \ref{fig:kyy} but for neglecting the Coulomb
modification. \label{fig:woc}}
\end{figure}

FIG. \ref{fig:kyy} shows our solution of $K_y^{y'}$
together with the data in Ref.~\cite{Kata}.
They are almost covering the error bar at $\theta_{\rm c.m.} = 0^\circ$,
thus the experimental data is described with the FY equation.
The calculations are numerically converged enough for this discussion.
The difference is about 2\% between the present results and
approximated ones including only $^1$S$_0$, $^3$S$_1$-$^3$D$_1$ states
in the two-body subsystem.
We expect that the influence to the results
is small with inclusion of the higher spin states.

Once we reported similar calculations in Ref.~\cite{uzu4,uzu5},
however, the graphs for the $K_y^{y'}$ show
our results without the Coulomb modification.
Then the discussions in the papers were based on these wrong results.

We found two discrepancies between our results and those in Ref.~\cite{Kata}.
1) Ours are sensitive to the incident deuteron energy
at around $\theta_{\rm c.m.} = 0^\circ$,
where the larger the energy becomes, the smaller the value is.
Then we expect if the target is as thin as the beam going through,
the measured data will be small.
2) There are peaks at around $\theta_{\rm c.m.} = 60^\circ$-$90^\circ$.
We suggest to measure at around those angles.

\begin{figure}[!t]
\includegraphics[width=80mm]{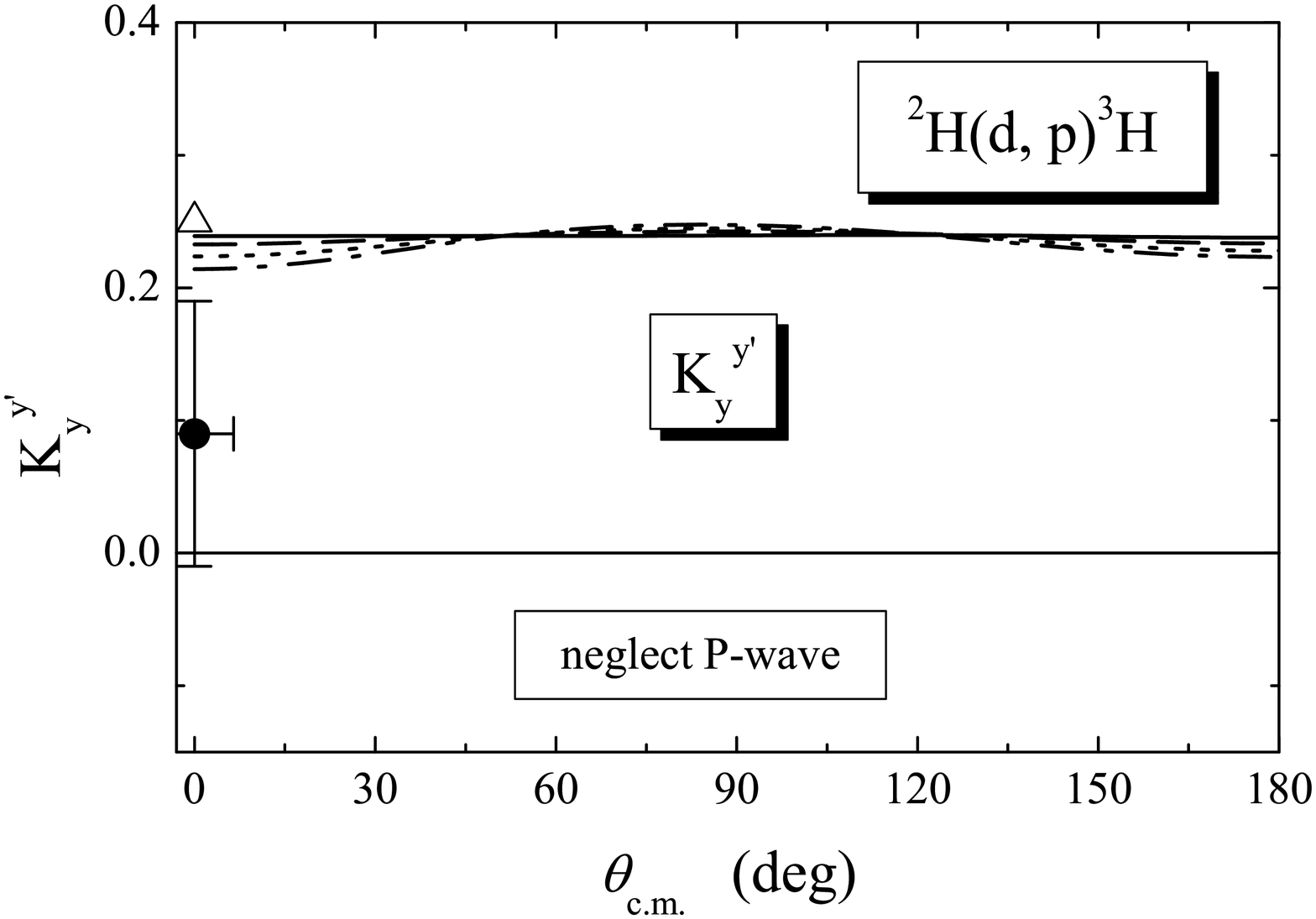}
\caption{Same as for FIG. \ref{fig:kyy} but for disregard of the P-wave components
in the initial channel. \label{fig:neglP}}
\end{figure}
\begin{figure}[!t]
\includegraphics[width=80mm]{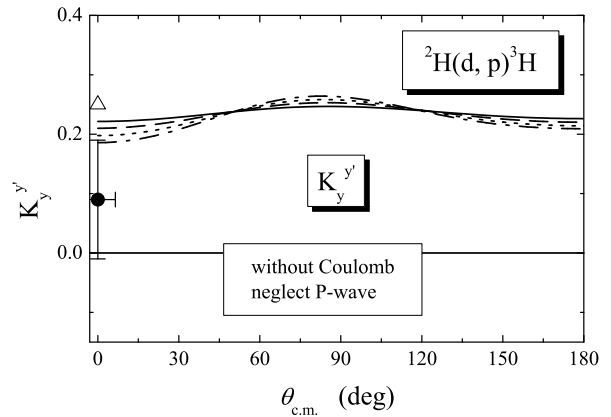}
\caption{Same as for FIG. \ref{fig:kyy} but for neglecting both of the P-wave
components in the initial channel and Coulomb modification. \label{fig:wocneglP}}
\end{figure}

To find out an origin of the sensitivity to the energy at $\theta_{\rm c.m.} = 0^\circ$,
we compare the results for without Coulomb modification (FIG. \ref{fig:woc})
with the original (FIG. \ref{fig:kyy}).
The former case $K_y^{y'}$ is negative at around $\theta_{\rm c.m.} = 0^\circ$
and positive at around $\theta_{\rm c.m.} = 180^\circ$,
which is totally different from the latter.
And the difference between $E_{\rm lab.} = 30$keV and 90keV for the former
is smaller than the latter by about 1/3 at $\theta_{\rm c.m.} = 0^\circ$.
Thus we think that most of the energy dependence is due to the Coulomb force.
This is against to the numerical results in Ref.~\cite{Kata}
since the energy variation of their amplitudes depends only on
the Coulomb penetrability function.

We make another check on a connection between the energy dependence and partial waves.
Since S- and P-waves are the main components in the initial channel
of low energy $^2$H$(d,p)^3$H reaction, $K_y^{y'}$ becomes almost isotropic
by disregard of the P-wave components from the reaction amplitudes (FIG.\ref{fig:neglP}).
In this case the Coulomb effects are canceled within our modification
by multiplication of an amplitude and one's complex conjugation.
Therefore the difference is small between FIG. \ref{fig:neglP}
and \ref{fig:wocneglP} which is neglecting both of the P-waves and Coulomb.
Then we find with comparison FIG. \ref{fig:kyy} - \ref{fig:neglP}
that the energy dependence in FIG. \ref{fig:kyy} comes from the Coulomb
modification which affects to the mixing term of S- and P-wave components in
the initical channel.

In case for neglecting P-waves, energy dependence at $\theta_{\rm c.m.} = 0^\circ$
is small and our solutions are close to that by Katabuch et al. using
the amplitudes in ref.~\cite{LSsup}.
It is interesting to us and further investigation is in progress.

\vspace{1\baselineskip}

The author appreciates helpful discussions with professors S. Oryu
and M. Tanifuji.
The calculations for this article are carried out on computers in
Research Center for Nuclear Physics, National Institute for Fusion
Science, The Institute of Physical and Chemical Research (RIKEN),
and Frontier Research Center for Computational Science in
Tokyo University of Science.

\end{document}